\documentclass[12pt,preprint]{aastex}
\usepackage{graphicx}
\usepackage{times,psfig}

\def\ltsima{$\; \buildrel < \over \sim \;$}
\def\lsim{\lower.5ex\hbox{\ltsima}}
\def\gtsima{$\; \buildrel > \over \sim \;$}
\def\gsim{\lower.5ex\hbox{\gtsima}}

\newcommand{\be}{\begin{equation}}
\newcommand{\en}{\end{equation}}
\newcommand{\ergs}{\rm \ erg \; s^{-1}}

\def\cmdue {\rm \ cm^{-2}}

\def\msole {~M_{\odot}}

\begin{document}

\title{Swift observations of SAX J1808.4--3658: monitoring the return 
to quiescence}  

\author{Sergio Campana\altaffilmark{1}, Luigi Stella\altaffilmark{2}, Jamie
A. Kennea\altaffilmark{3}}

\altaffiltext{1}{INAF-Osservatorio Astronomico di Brera, Via Bianchi
46, I--23807 Merate (Lc), Italy}
\altaffiltext{2}{INAF-Osservatorio Astronomico di Roma, Via di
Frascati 33, I-00040 Monteporzio Catone (Roma), Italy}
\altaffiltext{3}{Department of Astronomy and Astrophysics, 525 Davey
Lab., Pennsylvania State University, University Park PA 16802, USA}

\authoremail{sergio.campana@brera.inaf.it}


\begin{abstract}

The transient accreting millisecond pulsar SAX J1808.4--3658 has shown several outbursts to date but the transition from outburst to quiescence has never been investigated in detail. Thanks to the Swift observing flexibility, we monitored for the first time the decay to quiescence during the 2005 outburst. At variance with other transients, wide luminosity variations are observed. In addition, close to quiescence, SAX J1808.4--3658 seems to switch between two different states. We interpret them in terms of the accretion states accessible to a magnetized, fast rotating neutron star.

\end{abstract}


\section{Introduction}

SAX J1808.4--3658 (SAX J1808 in the following) was the first neutron star
transient during whose outbursts coherent pulsations were detected (Wijnands \& van der
Klis 1998). This source, together with six other low mass transients (see
Wijnands 2005 for a review), form a separate (sub-)class within neutron star
transients (see Campana et al. 1998a for a review). 
Distinctive properties, besides coherent pulsations, are weaker outburst peak
luminosities ($\sim (1-5)\times 10^{36}\ergs$), very low mass companions (mass
functions $<2\times 10^{-3}\msole$), short orbital periods ($P_{\rm orb}\lsim 4.3$ hr),
very faint quiescent luminosities ($\lsim 5\times 10^{31}\ergs$) and absence of
a soft X--ray spectral component in quiescence.

A further difference with respect to `classical' neutron star transients
concerns the return to quiescence following an outburst. The best light curve
to date of `classical' neutron star transients  
is represented by BeppoSAX observations of Aql X-1 (Campana et al. 1998b). 
Below a luminosity level of $\sim 10^{36}\ergs$ Aql X-1 turned off with an
exponential decay with an $e-$folding time of $\sim 1$ d. The source then
remained quiescent for at least one month following the 1998 outburst (Campana
et al. 1998b) and for five months following the 2000 outburst (Rutledge et
al. 2001; Campana \& Stella 2003). The return to quiescence for transient
pulsating neutron stars has been monitored for SAX J1808 (Wijnands et
al. 2003; Wijnands 2005) and for XTE J1751--305 (Markwardt et al. 2002) and
Swift J1756.9--2508 (Krimm et al. 2007). The behaviour of the latter two
sources was  somewhat
similar to Aql X-1 with a fast decay, even though in the case of XTE J1751--305
there was still activity 15 d after the transition to quiescence had occurred
(according to RXTE/PCA, corresponding to a 0.5--10 keV unabsorbed luminosity
of $\sim 10^{35}\ergs$ for a source distance of 7 kpc). 
In case of SAX J1808 the decay was completely different from that
observed in Aql X-1. The most striking feature is the strong erratic
variability by a factor of $\gsim 30$. This erratic behaviour was
not concentrated in the first few weeks after turn off but lasted for months
(e.g. Wjinands et al. 2004). The most likely explanation for this behaviour
comes from the similarities with dwarf novae where a similar behaviour has
been observed (Kato et al. 2004; Patterson et al. 2002) and to a lesser extent in
black hole transients (Kuulkers, Howell \& van Paradijs 1996). In the case of
dwarf novae this has been interpreted as due to a combination of short orbital
period and low mass ratio (Osaki \& Meyer 2002; Truss et al. 2002; Campana et
al. 2008, in preparation). 

Here we report on a monitoring campaign carried out with Swift, following the 2005
outburst. At variance with previous monitoring with RXTE/PCA we are able to
follow the source down to very low levels, a factor of $\sim 50$ fainter than
before, close to quiescence. In Sect. 2 we discuss the data and their analysis. Sect. 3
contains our discussion and conclusions.

\section{Data analysis}

Swift carried out 23 observations of SAX J1808 between
Jun 17, 2005 and Oct 28, 2005. XRT collected a total of 4899 s data in
Window Timing (WT) mode and 72922 s data in Photon Counting mode (see
Table 1). In WT mode a 1D image is obtained reading data compressing
along the central 200 pixels in a single raw. PC mode produces
standard 2D images (for more details see Hill et al. 2004). WT data
were mainly collected during the early stages of the outburst when the
source was brighter. PC data collected during this period are
piled-up.  

\begin{table*}
\caption{Observation log.}
\begin{center}
\begin{tabular}{cccccc}
Obs. ID.    & Date    &WT exp. time&PC exp. time& Count rate & Counts\\
            & (2006)  &  (s)       &  (s)       &(c s$^{-1}$)&       \\
00030034001 & 17 Jun  &    119     & 787        &  4.46      & 3510 (P)\\
00030034002 & 20 Jun  &    0       & 1061       &  1.49      & 1581 (P)\\
00030034003 & 23 Jun  &    761     & 206        &  3.01      &  620 (P)\\
00030075001 & 29 Jun  &    1043    & 0          &  4.09      & 4396    \\
00030034005 & 07 Jul  &    866     & 0          &  8.14      & 7157    \\
00030034006 & 13 Jul  &    733     & 394        &  3.10      & 1221 (P)\\
00030034010 & 02 Aug  &    0       & 416        &  8.88E-03  &    4 [A]\\
00030034011 & 05 Aug  &    0       & 1041       &  6.31E-03  &    7 [A]\\
00030034012 & 13 Aug  &    1028    & 2486       &  1.84      & 4574 (P)\\
00030034013 & 21 Aug  &    0       & 1675       &  9.05E-03  &   15 [A]\\
00030034014 & 28 Aug  &    68      & 1928       &  1.36      & 2622 (P)\\
00030034015 & 30 Aug  &    0       & 2550       &  6.26E-02  &  160 \\
00030034016 & 31 Aug  &    0       & 2838       &  9.87E-03  &   28  [A]\\
00030034017 & 02 Sep  &    0       & 2084       &  5.48E-03  &   11  [A]\\
00030034018 & 14 Sep  &    0       & 9011       &  1.50E-03  &   14  [B]\\
00030034019 & 25 Sep  &    123     & 7329       &  6.47E-03  &   47  [A]\\
00030075020 & 30 Sep  &    42      & 4922       &  4.89E-03  &   21  [A] \\
00030034021 & 04 Oct  &    0       & 7757       &  1.74E-03  &   12  [B]\\
00030034020 & 11 Oct  &    0       & 2986       &  4.89E-03  &   15  [A]\\
00030034022 & 12 Oct  &    0       & 4689       &  1.64E-03  &    7  [B]\\
00030034023 & 16 Oct  &    0       & 6742       &  1.83E-03  &   12  [B]\\
00030034024 & 20 Oct  &    0       & 4736       & $<$3.49E-03&   16  \\
00030034025 & 28 Oct  &    116     & 7284       & $<$3.00E-03&   22  \\
\end{tabular}
\end{center}

Labels in the last column indicate: (P) piled-up photon counting data; [A]
state A observation; [B] state B observation.
\end{table*}

The XRT data were processed with standard procedures (xrtpipeline ver. 0.11.6
within FTOOLS in the Heasoft package ver. 6.4), filtering, screening, and
grade selection criteria (Burrows et al. 2005). For the WT data,
we extracted source events in a square region with a side of 20
pixels. Ancillary response files generated with {\tt xrtmkarf} and accounted for
different extraction regions, vignetting, and point-spread function (PSF)
corrections. In PC mode we extracted data with variable extraction regions
depending on source strength, ranging from annular region with inner (outer)
radius of 10 (40) pixels when the source was very bright, down to 10 pixel
circular region when the source was very faint in order to increase the source
signal vs. the background.

\begin{figure*}[htbp]
\begin{center}
\psfig{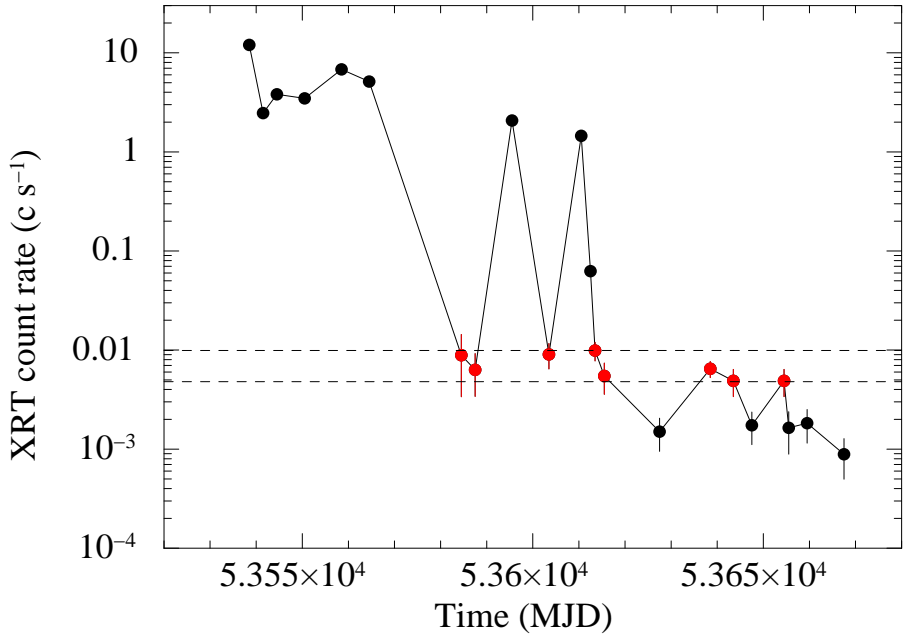}
\end{center}
\caption{SAX J1808 light curve observed with Swift XRT during the 2005 outburst.}
\label{lc}
\end{figure*}

\subsection{Outburst light curve}

The first observations caught SAX J1808 during the latest stages of the
outburst phase at a count rate level (corrected for pile-up) of more than 10
counts s$^{-1}$. Starting from MJD 53564 the source decreased its count rate by
a factor of $\sim 600$ in 20 days (unfortunately no observations were taken
during this period due to the occurrence of number of gamma--ray bursts, see Fig. 1). Two
observations within 3 days showed the source at a level of 0.01 counts
s$^{-1}$ well above the quiescent level. After this sharp decay, SAX J1808
entered a ``flaring'' behaviour (Wijnands et al. 2003; Wijnands 2005). Our data
evidence that this activity lasted for about one month; after that the source returned
to its quiescent state with some comparatively small scale variability (factor of $\sim 5$)
superimposed (see also below). 

\begin{figure*}[!htbp]
\begin{center}
\psfig{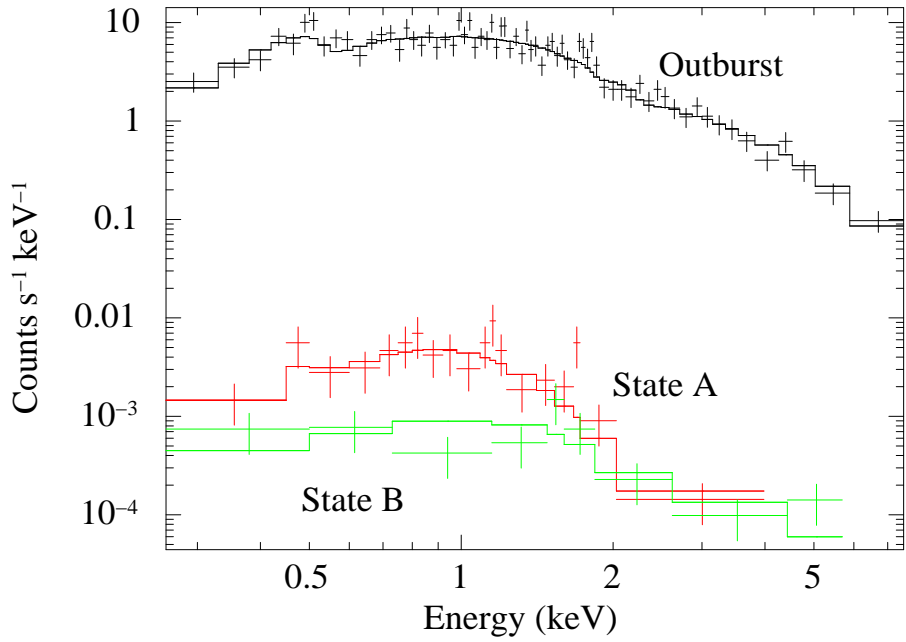}
\end{center}
\caption{{XRT spectra of SAX J1808.4--3658. The topmost spectrum is a representative of 
the outburst state. The middle spectrum refers to state A and the lowest one to state B (see text
for more details). All spectra were fitted with an absorbed black body plus power law model.}}
\label{totspe}
\end{figure*}

\subsection{Spectral analysis: bright end}

We fit the entire data set with an absorbed black body plus power law model,
keeping the same value of the column density for all the spectra. The best column density
is $(1.3\pm0.1)\times 10^{21}\cmdue$ ($90\%$ confidence level, modelled with
{\tt TBABS}). In Table 2 we report the results of our spectral modelling. From
this table and Figs. 2 \& 3 it is clear that during the outburst decay and flaring
period the power law photon index ($\Gamma\sim 2.3$) and the black body
temperature ($k\,T\sim 0.65$ keV) remained almost constant. This behaviour
occurs in the (unabsorbed) 0.3--10 keV luminosity range of
$7\times10^{34}-10^{36}\ergs$ (i.e. a factor of $\sim 15$), assuming a source distance of 3.5
kpc (Galloway \& Cummings 2006). 

\begin{table*}
\caption{Spectral fits.}
\begin{tabular}{ccccc}
Date          &BB Temperature   &  BB Radius     & PL index & Luminosity\tablenotemark{a} \\
              & (keV)           &   (km)         &          & (erg s$^{-1}$)      \\
2005-06-17\tablenotemark{b}&$0.72^{+0.13}_{-0.13}$&$2.3^{+1.0}_{-0.9}$ &$2.24^{+0.15}_{-0.11}$& $1.0\times10^{36}$\\
2005-06-20\tablenotemark{b}&$0.80^{+0.17}_{-0.14}$&$1.2^{+0.6}_{-0.6}$ &$2.51^{+0.41}_{-0.38}$& $2.4\times10^{35}$\\
2005-06-23\tablenotemark{b}&$0.59^{+0.15}_{-0.13}$&$2.4^{+1.6}_{-1.6}$ &$2.26^{+0.11}_{-0.08}$& $7.2\times10^{35}$\\
2005-06-29\tablenotemark{b}&$0.78^{+0.19}_{-0.19}$&$1.0^{+0.4}_{-0.4}$ &$2.37^{+0.15}_{-0.12}$& $3.3\times10^{35}$\\
2005-07-07\tablenotemark{b}&$0.44^{+0.23}_{-0.33}$&$2.2^{+2.2}_{-1.9}$ &$2.41^{+0.07}_{-0.03}$& $5.7\times10^{35}$\\
2005-07-13\tablenotemark{b}&$0.57^{+0.10}_{-0.09}$&$2.2^{+0.9}_{-0.8}$ &$2.18^{+0.11}_{-0.08}$& $4.2\times10^{35}$\\
2005-08-13\tablenotemark{b}&$0.68^{+0.24}_{-0.19}$&$0.9^{+0.6}_{-0.5}$ &$2.22^{+0.12}_{-0.09}$& $2.0\times10^{35}$\\
2005-08-28\tablenotemark{b}&$0.78^{+0.18}_{-0.17}$&$0.8^{+0.4}_{-0.4}$ &$1.94^{+0.16}_{-0.05}$& $1.4\times10^{35}$\\
\hline
State A\tablenotemark{c}  &$0.20^{+0.05}_{-0.05}$&$1.3^{+0.3}_{-0.5}$ &1.93 fixed       & $5.0\times10^{32}$\\
State B\tablenotemark{d}  &0.20 fixed       &$<0.3$          &1.93 fixed       & $1.5\times10^{32}$\\
\end{tabular}

\noindent \tablenotetext{a}{Unabsorbed 0.3--10 keV luminosity at a source distance of 3.5 kpc.}

\noindent \tablenotetext{b}{Overall reduced $\chi^2=1.06$ with 1213 degrees of freedom.}

\noindent \tablenotetext{c}{Reduced $\chi^2=0.75$ with 16 degrees of freedom.}

\noindent \tablenotetext{d}{Reduced $\chi^2=1.38$ with 7 degrees of freedom.}

\end{table*}

\subsection{Spectral analysis: faint end}

At lower luminosities the number of collected photons is very small and we have
to stack together different observations. One important consideration comes
from the observation that 8 out of 13 observations found the source with a
count rate in the narrow interval $5-10$ c ks$^{-1}$, whereas the other
5 all lie within $0.8-2$ c ks$^{-1}$. The mean of the first group is
$6.3\pm1.1$ c ks$^{-1}$ and of 
the second is $1.3\pm0.4$ c ks$^{-1}$. This is rather peculiar and hints for
the existence of two different states in the deep faint end tail of the outburst.
Due to the small number of photons these spectra can be easily fit with single
power law models. For state A we derive a photon index of $\Gamma=2.7\pm0.3$
and a mean unabsorbed 0.3--10 keV luminosity of $5.0\times 10^{32}\ergs$, for
state B we have $\Gamma=1.7\pm0.6$ and a luminosity of $1.5\times
10^{32}\ergs$. These values have been derived binning the data to 5 photons
per energy bin and applying the Churazov weighting in the fitting procedure 
(see Fig. 2).
This indicates that state B is not yet the true quiescent state, characterized
by a single power law spectrum with index $\Gamma=1.93^{+0.37}_{-0.29}$ and a
source luminosity of $(5.2\pm0.1)\times 10^{31}\ergs$ (Heinke et al. 2007),
i.e. a factor of $\sim 3$ smaller that state B luminosity.
Assuming this power law component as a stable component (at least in photon
index) present in state A and B spectra, one can investigate if a black body
component is present. Fitting state A spectrum with a fixed power law photon
index (with $\Gamma=1.93$) and a free black body component, we
derive $k\,T=0.20\pm0.05$ keV and a radius of $R=1.3^{+0.3}_{-0.5}$
km. The improvement over the simple power law fit is at a level
of $2.2\,\sigma$ by means of an F-test. For state B we are not able
to constrain the black body component. However, if we assume the same
black body temperature as in state A (i.e. 0.2 keV) we can derive an
upper limit on its radius of $R<0.3$ km, indicating that if present
and at the same temperature it must be smaller in size. 

\begin{figure*}[!htbp]
\begin{center}
\psfig{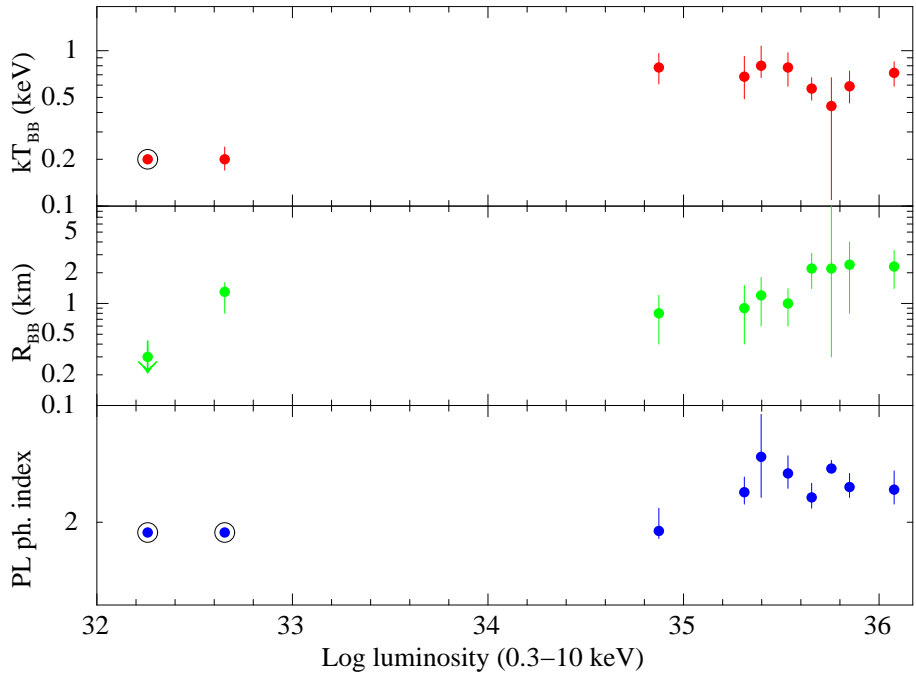}
\end{center}
\caption{Spectral parameters evolution across the outburst. Circled values
indicate a fixed parameter in the fitting procedure.}
\label{para}
\end{figure*}

\section{Discussion and conclusions}

SAX J1808 is the prototype of the accreting millisecond X--ray pulsar (AMP)
class. The source underwent several outbursts to date which have been
monitored in great details thanks to RXTE observations (Wijnands 2005;
Wijnands et al. 2003). These observations however were limited to the
brightest part of the outburst, being the PCA on RXTE a collimated instrument
(and therefore heavily background limited). Here we report on the first
campaign aimed at studying the faintest tail of the outburst and return to
quiescence. Thanks to the fast repointing and flexible scheduling capabilities
of the Swift satellite,
we monitored the return to quiescence of SAX J1808 during the 2006 outburst.
The bright phase of the outburst is similar to what has been observed in the
past: following the peak and the smooth decay a flaring behaviour sets in with
at least three rebrightening episodes. After this the source turns to
quiescence even if same low level activity (factor of $\sim 5$) is still
present. 

The puzzling result of this observational campaign on SAX J1808 is that out of
13 observations with luminosity below $\sim 10^{34}\ergs$, 8 times the source
was found with a luminosity around $\sim 5\times 10^{32}\ergs$ (see
Fig. 1), i.e. at a luminosity level $\sim 10$ times higher than the true
quiescent level (Campana et al. 2002; Heinke et al. 2007). Given the wild
luminosity variations this is somewhat strange and calls for the presence of
a `metastable' state. This luminosity level is 
difficult to interpret in case of standard disk accretion but finds a natural
explanation if the accretion process is mediated by the magnetic field of the
fast spinning neutron star. When the magnetospheric radius (i.e. the radius at
which the neutron star magnetic field starts controlling the accretion flow)
is larger than the corotation radius, the system is in the (so-called)
propeller regime. In this case matter is halted from the rotating
magnetosphere and does not accrete onto the neutron star surface (see
e.g. Campana \& Stella 2000 and references therein). When the magnetospheric
radius becomes larger than the light cylinder radius ($r_{\rm
lc}=c\,P/2\,\pi$, where $c$ is the velocity of light and $P$ the spin period),
the magnetic field cannot corotate any more with the neutron star and a dipole
losses will take place. The luminosity range spanned in the propeller regime
is $\sim 440\,P_{\rm 2.5}^{3/2}\,M_{1.4}^{-3/2}$ (where $P_{\rm 2.5}$ is the
spin period in units of 2.5 ms and $M_{1.4}$ the neutron star mass in units of
$1.4\msole$; see Campana \& Stella 2000). In order to have the `metastable'
state to coincide with the lowest luminosity in the propeller regime
(i.e. just before the reactivation of the pulsar), we have to require a
magnetic field of $\sim 7\times 10^7$ G, in line with previous estimates: 
$\sim (1-10)\times 10^8$ G using disk-magnetosphere interaction models
(Psaltis \& Chakrabarty 1999), $\sim (1-6)\times 10^8$ G using simple
considerations on the position of the magnetospheric radius during quiescent
periods (Di Salvo \& Burderi 2003); $(0.4-1.5)\times 10^8$ G modelling period
changes during outbursts with a magnetic torque model (Hartman et
al. 2008).   

The lowest luminosity level in the proposed interpretation is ascribed to the
interaction of a turned-on pulsar with the interbinary and circumstellar
environment. Variability is expected at this stage due to the rapidly changing
environment as observed, e.g., in the millisecond radio pulsar PSR J1740--5340
(D'Amico et al. 2001; Ferrario et al. 2001) which gets eclipsed over a range
of orbital phases for different orbital cycles.

\end{document}